\documentclass[amssymb,floatfix,epj]{svjour}
\usepackage{amsmath}
\usepackage{mathtools,cuted}
\usepackage{graphicx}
\usepackage{amssymb}
\usepackage{siunitx}
\usepackage{dcolumn}
\usepackage{bm}
\usepackage{epsfig}
\usepackage{setspace}
\usepackage[caption=false]{subfig}
\usepackage{xcolor}
\usepackage[numbers]{natbib}
\begin{document}
\title{Broadening of the chiral critical region in a hydrodynamically expanding medium}
\author{Christoph Herold\inst{1} \and Marcus Bleicher\inst{2,4,5,6} \and Marlene Nahrgang\inst{3} \and Jan Steinheimer\inst{2} \and Ayut Limphirat\inst{1} \and Chinorat Kobdaj\inst{1} \and Yupeng Yan\inst{1}
}                     
\offprints{}          
\institute{School of Physics, Suranaree University of Technology, 111 University Avenue, Nakhon Ratchasima 30000, Thailand \and Frankfurt Institute for Advanced Studies (FIAS), Ruth-Moufang-Str.~1, 60438 Frankfurt am Main, Germany \and SUBATECH, UMR 6457, IMT Atlantique, Universit\'{e} de Nantes, IN2P3/CNRS, 4 rue Alfred Kastler, 44307 Nantes cedex 3, France \and GSI Helmholtzzentrum f\"ur Schwerionenforschung GmbH, Planckstr. 1, 64291 Darmstadt, Germany \and John von Neumann-Institut f\"ur Computing, Forschungzentrum J\"ulich, 52425 J\"ulich, Germany \and Institute for Theoretical Physics, Goethe-University, 60438 Frankfurt, Germany}
\date{Received: date / Revised version: date}
%
\abstract{
We investigate higher cumulants of the sigma field as the chiral order parameter at the QCD phase transition. We derive a thermodynamic expression for the skewness and kurtosis from susceptibilities and use these to determine $S\sigma$ and $\kappa\sigma^2$ for the sigma field in equilibrium. In a next step, we study the behavior of these cumulants in both an equilibrated static medium and an expanding medium resembling the hydrodynamic stage of a heavy-ion collision. For the latter one, we find a significant broadening of the critical region. 
\PACS{
      {PACS-key}{discribing text of that key}   \and
      {PACS-key}{discribing text of that key}
     } 
} 
\maketitle
\section{Introduction}

Under extreme conditions, such as large temperatures and densities, a new phase of strongly-interacting matter exists which is characterized by the restoration of chiral symmetry and the deconfinement of color charges. Evidence for such a quark-gluon plasma (QGP) has been drawn from results at ultrarelativistic heavy-ion collision at RHIC and LHC. The transition from a hadronic medium to a QGP has been discovered by lattice QCD to be an analytic crossover rather than a phase transition \cite{Aoki:2006we,Borsanyi:2010bp,Borsanyi:2013bia,Bazavov:2014pvz}. This, however, is true only for the case of vanishing or small baryochemical potential $\mu_{\rm B}$, as the fermionic sign problem complicates standard lattice QCD methods for large densities. In this regime, new methods have to be developed such as a systematical Taylor expansion of the pressure in terms of $\mu_{\rm B}/T$ \cite{Fodor:2001au,deForcrand:2002ci,Endrodi:2011gv}. From this, the existence of a QCD critical point has been ruled out up to values of $\mu_{\rm B}/T\lesssim1$.

Applying functional methods, a critical point (CP) and first-order phase transition in the regime of large $\mu_{\rm B}$ have been predicted \cite{Fischer:2014ata, Eichmann:2015kfa}. In this approach which does not suffer from the limitations of lattice QCD, the authors have studied a coupled set of Dyson-Schwinger equations for the dressed quark and gluon propagators. 

Experimentally, physicists hope to detect signals of a CP in heavy-ion collisions. Here, especially fluctuations of conserved charges are of interest and have been reported in the beam-energy scan program of RHIC \cite{Aggarwal:2010wy,Adamczyk:2013dal}. The measured data of the net-proton skewness and especially the kurtosis significantly deviates from baseline calculations with UrQMD or a hadron resonance gas model. However, great care is required to properly interpret the data. On the one hand, nonmonotonic behavior in susceptibilities and higher-order cumulants is expected near a CP where the correlation length diverges \cite{Asakawa:2000wh,Stephanov:1998dy,Stephanov:1999zu,Cheng:2008zh,Asakawa:2009aj,Gupta:2011wh,Friman:2011pf,Stephanov:2008qz,Athanasiou:2010kw}, on the other hand, these predictions rely on an equilibrated medium and neglect effects of the highly dynamical environment created in a heavy-ion collision.  

In nonequilibrium, a finite equilibration time has to be taken into account. Especially in the vicinity of a CP where its growth is proportional to a certain power of the correlation length determined by the dynamical universality class \cite{HALPERIN:1969zza}, critical slowing down severly limits the divergence of fluctuations. This has been studied phenomenologically in \cite{Berdnikov:1999ph}, where two important consequences have been found. First, the correlation length grows only up to values of $1.5-2$~fm, and second, the system remains correlated longer than it would in equilibrium. The latter is often called a memory effect. Its importance has been emphasized in \cite{Mukherjee:2015swa}, where the real-time evolution of non-Gaussian cumulants in a homogeneously expanding medium has been studied, showing a significant dependence on the thermalization time. This model, however, oversimplifies the situation found in experiment by neither taking into account a realistically expanding medium nor the proper interplay of fluctuations in the chiral order parameter with the surrounding heat bath. 

In the work presented here, we use the model of nonequilibrium chiral fluid dynamics (N$\chi$FD) to address the problem of the evolution of the chiral order parameter and its cumulants in a realistic setup. The sigma field as the chiral order parameter is explicitly propagated using a Langevin equation, taking into account its interaction with a fermi-onic heat bath to describe the fluid dynamical expansion of the hot and dense medium \cite{Nahrgang:2011mg}. We have demonstrated previously that this model captures essential features of the QCD chiral phase transition such as critical slowing down near a CP and spinodal decomposition for a first-order phase transition scenario \cite{Herold:2013qda,Nahrgang:2011mv,Herold:2013bi,Herold:2014zoa}. The latter one has also been investigated \cite{Mishustin:1998eq,Sasaki:2007db,Randrup:2009gp,Randrup:2010ax,Steinheimer:2012gc}. We were furthermore able to show how the net-proton kurtosis closely follows the kurtosis of the sigma field during a crossover transition \cite{Herold:2016uvv} left of the CP where a negative kurtosis of the sigma field is expected to yield a suppressed net-baryon and net-proton kurtosis \cite{Stephanov:2011pb}. We now focus on the evolution of the second, third, and fourth cumulant of the sigma field along hypersurfaces of constant energy density. We study their evolution in an isothermal box with linearly decreasing temperature and in a realistically expanding fluid. For both scenarios we compare the dynamically extracted cumulants to the corresponding temperature and chemical potential dependent quantities that are obtained from susceptibilities. Our results are especially interesting for experiments at the upcoming facilities of FAIR \cite{Friman:2011zz} and NICA \cite{nica:whitepaper}, or RHIC's BES II program. 
\begin{figure}[t]
{
\centering
    \includegraphics[width=0.5\textwidth]{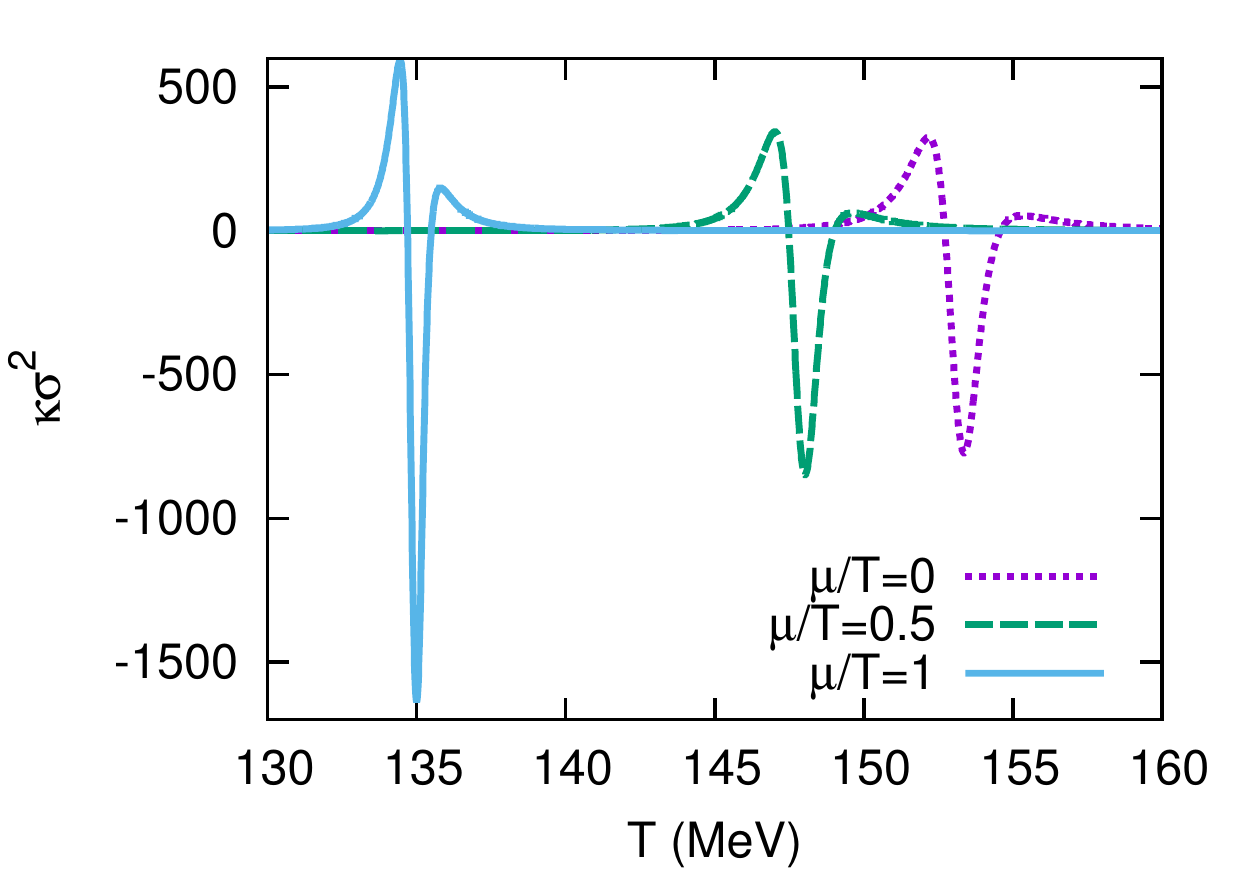}
}
\caption[kurtosis]{Kurtosis of the chiral order parameter for different values of $\mu/T$.}
\label{fig:kurt}
\end{figure}

Before presenting the numerical results, we give a general derivation of the higher order cumulants of a scalar field in Sec.~\ref{sec:eval}. That allows us to calculate the thermodynamic expectation values for these quantities. The results are then applied for illustrative purposes to a quark-meson model in Sec.~\ref{sec:qm}. We continue with a description of the N$\chi$FD model in Sec.~\ref{sec:nxfd}, followed by results of the calculations in a box and an expanding fluid in Secs.~\ref{sec:box} and \ref{sec:exp}. We summarize our observations and give a brief outlook in Sec.~\ref{sec:sum}.

\section{Evaluation of sigma fluctuations}
\label{sec:eval}

Our goal is to determine higher order cumulants of the sigma field in an effective model using derivatives of the effective potential. These quantities were identified as influencing experimental observables stemming from the critical region of the QCD phase diagram such as the net-proton fluctuations \cite{Stephanov:2011pb}. 

We start with the generating functional for connected diagrams in Euclidean space-time:
\begin{equation}
 W[J]=-\log Z[J]
\end{equation}
for a scalar field $\phi$ with source $J$ and generating functional
\begin{equation}
 Z[J]=\int {\mathcal D}\phi \mathrm e^{-S_E-\int J\phi}~,
\end{equation}
where $S_E$ denotes the action in Euclidean space-time, $S_E=\int_0^{1/T}\mathrm d\tau\int\mathrm d^3 x {\mathcal L}_E(x,\tau)$. 
We obtain the expectation value of $\phi$ in the standard fashion as
\begin{align}
 \langle \phi(x)\rangle=&\left.\frac{\delta W[J]}{\delta J(x)}\right|_{J=0}=\left.-\frac{1}{Z}\frac{\delta Z}{\delta J(x)}\right|_{J=0}\nonumber\\=&\frac{1}{Z}\int {\mathcal D}\phi\, \phi(x) \mathrm e^{-S_E}~.
\end{align}
Next, we determine higher derivatives of $W[J]$ as n-point functions. Starting with the second order, we get
\begin{align}
 \frac{\delta^2 W}{\delta J(x_1)\delta J(x_2)}=&\frac{\delta}{\delta J(x_2)}\left[-\frac{1}{Z}\frac{\delta Z}{\delta J(x_2)}\right]\nonumber\\=&\frac{1}{Z^2}\frac{\delta Z}{\delta J(x_1)}\frac{\delta Z}{\delta J(x_2)}-\frac{1}{Z}\frac{\delta^2 Z}{\delta J(x_1)\delta J(x_2)}\nonumber \\
 =&-\langle\phi(x_1)\phi(x_2)\rangle + \langle\phi(x_1)\rangle\langle\phi(x_2)\rangle~.
\end{align}
For $x_1=x_2=x$ and with $\phi-\langle\phi\rangle=\delta\phi$ this gives us
\begin{equation}
\label{eq:w2}
  \left.\frac{\delta^2 W}{\delta^2 J(x)}\right|_{J=0}=-\langle\delta\phi^2\rangle~.
\end{equation}
We continue now with higher derivatives. To third order we end up with
\begin{align}
 \frac{\delta^3 W}{\delta J(x_1)\delta J(x_2)\delta J(x_3)}=&-\frac{2}{Z^3}\frac{\delta Z}{\delta J(x_1)}\frac{\delta Z}{\delta J(x_2)}\frac{\delta Z}{\delta J(x_3)}\nonumber\\&+\frac{1}{Z^2}\frac{\delta Z}{\delta J(x_3)}\frac{\delta^2 Z}{\delta J(x_1)\delta J(x_2)}\nonumber\\ 
 &+\frac{1}{Z^2}\frac{\delta Z}{\delta J(x_1)}\frac{\delta^2 Z}{\delta J(x_2)\delta J(x_3)}\nonumber\\&+\frac{1}{Z^2}\frac{\delta Z}{\delta J(x_2)}\frac{\delta^2 Z}{\delta J(x_1)\delta J(x_3)}\nonumber\\
 &-\frac{1}{Z}\frac{\delta^3 Z}{\delta J(x_1)\delta J(x_2)\delta J(x_3)}~,
\end{align}
and
\begin{equation}
  \left.\frac{\delta^3 W}{\delta^3 J(x)}\right|_{J=0}=2\langle\phi\rangle^3-3\langle\phi\rangle\langle\phi^2\rangle+\langle\phi^3\rangle=\langle\delta\phi^3\rangle~.
\end{equation}
To determine the kurtosis, we need to go one more step and finally obtain
\begin{align}
  \left.\frac{\delta^4 W}{\delta^4 J(x)}\right|_{J=0}=&-\langle\phi^4\rangle-12\langle\phi\rangle^2\langle\phi^2\rangle+4\langle\phi\rangle\langle\phi^3\rangle\nonumber\\&+3\langle\phi^2\rangle^2+6\langle\phi\rangle^4\nonumber\\=&-\left[\langle\delta\phi^4\rangle-3\langle\delta\phi^2\rangle^2\right]~,
\end{align}
which resembles the negative kurtosis $-\kappa\langle\delta\phi^2\rangle$ of fluctuations in the field $\phi$. 

Our next goal is to express the moments in terms of derivatives of the one-particle (1PI) irreducible effective action with respect to the (classical) field $\phi$. It is defined as the Legendre transformation of the functional $W[J]$: 
\begin{equation}
 \Gamma[\phi]=W[J]-\int J\phi~.
\end{equation}
The derivative with respect to the classical field then yields the source $J$:
\begin{equation}
 \frac{\delta\Gamma[\phi]}{\delta\phi(x)}=-J(x)~.
\end{equation}
We now evaluate
\begin{align}
 &\int\mathrm d^4 z\frac{\delta^2 W}{\delta J(x_1)\delta J(z)}\frac{\delta^2 \Gamma}{\delta\phi(z)\delta\phi(x_2)}\nonumber\\
 =&\int\mathrm d^4 z\frac{\delta\phi(x_1)}{\delta J(z)}\left(-\frac{\delta J(z)}{\delta\phi(x_2)}\right)\nonumber\\
 =&-\delta(x_1-x_2)~.
\end{align}
Together with eq.~\eqref{eq:w2}, we can then determine the susceptibility as the inverse of the second derivative of the effective action, cf.~\cite{Sasaki:2006ww},
\begin{equation}
\label{eq:sus2}
 \langle\delta\phi^2\rangle=\left.-\frac{\delta^2 W}{\delta J(x)^2}\right|_{J=0}=\left(\frac{\delta^2\Gamma}{\delta\phi^2}\right)^{-1}~.
\end{equation}

Next, we derive a relation for the three-point function 
\begin{strip}
\begin{align}
\label{eq:w4}
 \frac{\delta^3 W}{\delta J(x_1)\delta J(x_2)\delta J(x_3)}=&\int\mathrm d^4 u\mathrm d^4 v\left(-\frac{\delta^2 W}{\delta J(x_1)\delta J(u)}\right)
 \frac{\delta}{\delta J(x_3)}\left(-\frac{\delta^2\Gamma}{\delta\phi(u)\delta\phi(v)}\right)\frac{\delta^2 W}{\delta J(v)\delta J(x_2)}\\
 =&\int\mathrm d^4 u\mathrm d^4 v\mathrm d^4 w\frac{\delta^2 W}{\delta J(x_1)\delta J(u)}\frac{\delta^2 W}{\delta J(x_2)\delta J(v)}\frac{\delta^2 W}{\delta J(x_3)\delta J(w)} 
 \frac{\delta^3\Gamma}{\delta\phi(u)\delta\phi(v)\delta\phi(w)}~. \nonumber
\end{align}
\end{strip}
In the first step we have used that 
\begin{equation}
\frac{\partial}{\partial\alpha}M^{-1}=-M^{-1}\frac{\partial M}{\partial\alpha}M^{-1}~,
\end{equation}
and in the second step the chain rule 
\begin{equation}
 \frac{\delta}{\delta J(x_3)}=\int\mathrm d^4 w \frac{\delta\phi(w)}{\delta J(x_3)}\frac{\delta}{\delta\phi(w)}~.
\end{equation}
We can then write down the third central moment of the field $\phi$ as
\begin{equation}
\label{eq:sus3}
 \langle\delta\phi^3\rangle=-\frac{\delta^3\Gamma}{\delta\phi^3}\left(\frac{\delta^2\Gamma}{\delta\phi^2}\right)^{-3}~.
\end{equation}
Differentiating once more with $\frac{\delta}{\delta J(x_4)}$ and using the same identities as in \eqref{eq:w4}, we can finally write down the expression for the kurtosis as
\begin{align}
\label{eq:sus4}
 \kappa\langle\delta\phi^2\rangle=&\left[\langle\delta\phi^4\rangle-3\langle\delta\phi^2\rangle^2\right]\nonumber\\=&-\frac{\delta^4\Gamma}{\delta\phi^4}\left(\frac{\delta^2\Gamma}{\delta\phi^2}\right)^{-4}+3\left(\frac{\delta^3\Gamma}{\delta\phi^3}\right)^2\left(\frac{\delta^2\Gamma}{\delta\phi^2}\right)^{-5}~.
\end{align}

\section{The chiral kurtosis in the quark-meson model}
\label{sec:qm}

As an example we calculate the kurtosis of the sigma field in the quark-meson model, given by the Lagrangian
\begin{align}
\label{eq:Lagrangian}
 {\cal L}&=\overline{q}\left(i \gamma^\mu \partial_\mu-g_{\rm q} \sigma\right)q + \frac{1}{2}\left(\partial_\mu\sigma\right)^2- U(\sigma)~, \\
 U(\sigma)&=\frac{\lambda^2}{4}\left(\sigma^2-f_{\pi}^2\right)^2-f_{\pi}m_{\pi}^2\sigma~.    
\end{align}
with the light quark doublet $q=(u,d)$ and the chiral condensate $\sigma$ which dynamically generates the mass of the constituent quarks. This Lagrangian is also used as the basis for the N$\chi$FD model.
We can fix the coupling constant $g$ from the vacuum nucleon masses to $g=3.37$. The additional parameters are the pion decay constant of $f_\pi=93$~MeV and the pion mass $m_\pi=138$~MeV. The term proportional to $\sigma$ accounts for the small explicit symmetry breaking due to the finite current quark masses. The self-coupling constant $\lambda$ is related to the sigma mass $m_\sigma=\SI{600}{\MeV}$ through $\lambda^2=\frac{m_\pi^2-m_\sigma^2}{2f_\pi^2}$. More details can be found in \cite{Sasaki:2011sd,Herold:2014zoa}. 

This model is well studied and we can immediately write down the mean-field effective thermodynamic potential as
\begin{strip}
\begin{align} 
\label{eq:thermpot}
\Omega(\sigma)&=\frac{T}{V}\Gamma[\sigma]=U(\sigma)-\Omega_{\rm q\bar q}(T,\mu; \sigma)~,\\
\Omega_{\rm q\bar q}(T,\mu; \sigma)&=
d_q T\int\frac{\mathrm d^3 p}{(2\pi)^3} \left\{\ln\left[1+\mathrm e^{-\frac{E_{\rm q}-\mu}{T}}\right]
+\ln\left[1+\mathrm e^{-\frac{E_{\rm q}+\mu}{T}}\right]\right\}~,
\end{align}
\end{strip}
with the degeneracy factor $d_q=24$ and the quasiparticle energy $E_{\rm q}=\sqrt{p^2+m_{\rm q}^2}$. Note that here the quark chemical potential with $\mu=\mu_{\rm B}/3$ is used. We then evaluate the kurtosis with the help of the derivatives according to eq.~\eqref{eq:sus4} and show the result in fig.~\ref{fig:kurt} for various values of $\mu/T$ on the crossover side left of the CP in the corresponding phase diagram. 

\begin{figure}[h!]
\centering
    \subfloat[\label{fig:varboxsig}]{
    \centering
    \includegraphics[scale=0.65,angle=270]{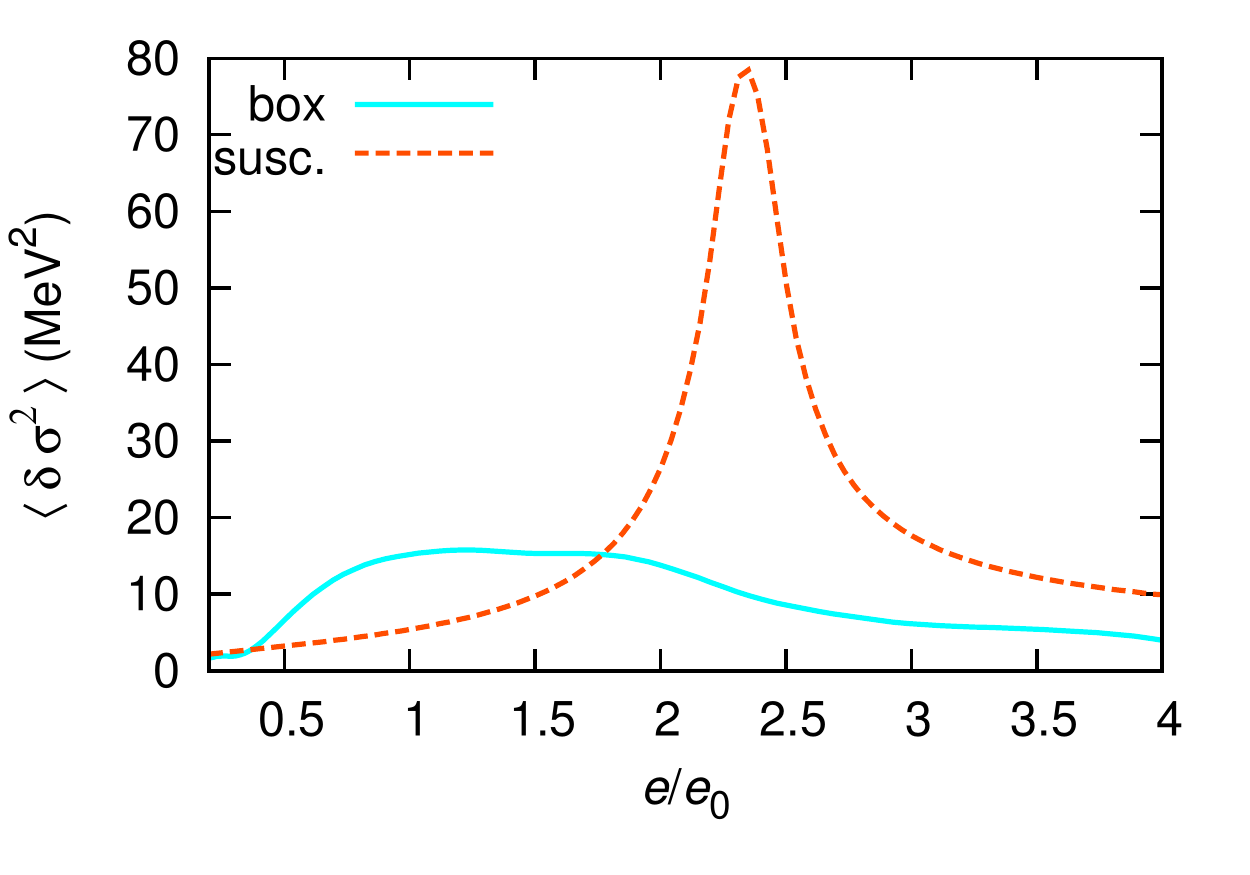}
    }   
  \vfill
    \subfloat[\label{fig:skewboxsig}]{
    \centering
    \includegraphics[scale=0.65,angle=270]{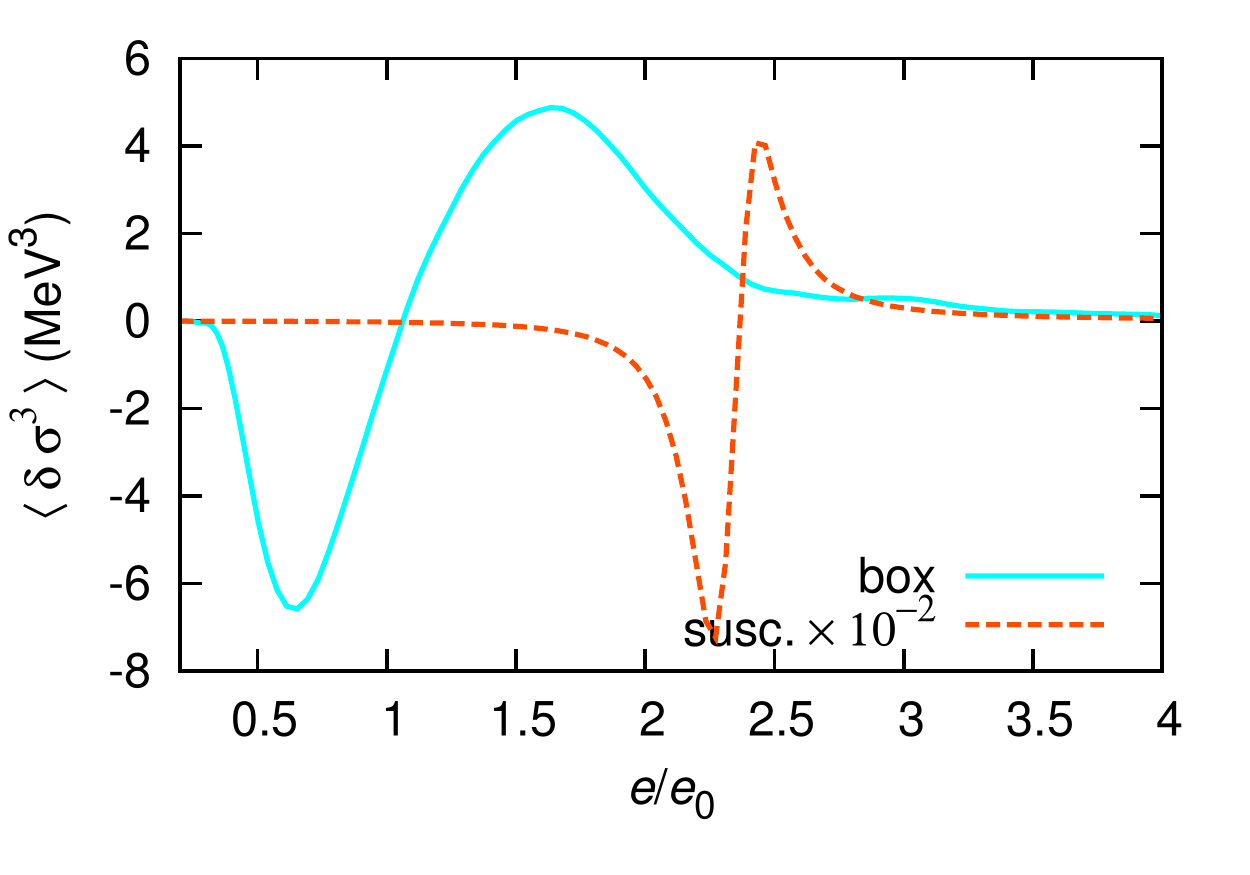}
    }
  \vfill
    \subfloat[\label{fig:kurtboxsig}]{
    \centering
    \includegraphics[scale=0.65,angle=270]{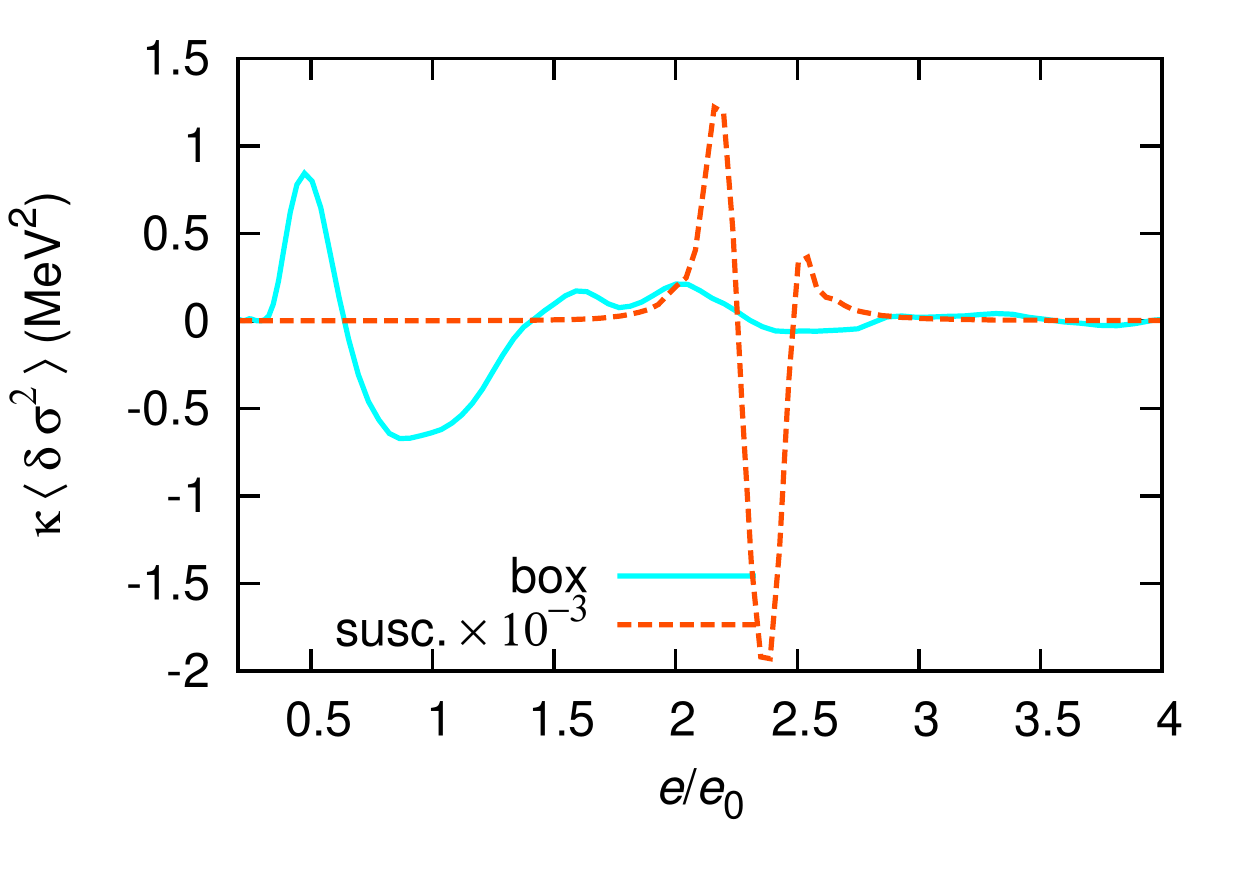}
    }  
\caption[maps]{Second \subref{fig:varboxsig}, third \subref{fig:skewboxsig}, and fourth \subref{fig:kurtboxsig} cumulants of the sigma field in a box with temperature linearly dropping from $\SI{160}{\MeV}$ to $\SI{80}{\MeV}$ (solid lines) compared to the equilibrium susceptibilities (dashed lines). We clearly see a memory effect in the evolution delaying the formation of dips and peaks in comparison to the equilibrium values.}
\label{fig:box}
\end{figure}

\section{Nonequilibrium chiral fluid dynamics}
\label{sec:nxfd}

The starting point of the N$\chi$FD model is the quark-meson Lagrangian from Eq.\ \eqref{eq:Lagrangian} with the thermodynamic potential Eq.\ \eqref{eq:thermpot}. In the full nonequilibrium dynamics, we explicitly propagate the chiral order parameter with a Langevin equation of motion, derived from the 2PI effective action \cite{Nahrgang:2011mg}
\begin{equation}
\label{eq:eomsigma}
 \partial_\mu\partial^\mu\sigma+\eta\partial_t \sigma+\frac{\delta \Omega}{\delta\sigma}=\xi~,
\end{equation}
The damping coefficient $\eta$ arises from the $\sigma\leftrightarrow q\bar q$ reaction and has been evaluated as
\begin{equation}
\label{eq:dampingcoeff}
  \eta=\frac{12 g^2}{\pi}\left[1-2n_{\rm F}\left(\frac{m_\sigma}{2}\right)\right]\frac{1}{m_\sigma^2}\left(\frac{m_\sigma^2}{4}-m_{\rm q}^2\right)^{3/2}~.
\end{equation}
The stochastic noise term $\xi$ is assumed to be Gaussian and white, its width determined by the dissipation-fluctuation relation
\begin{equation}
\label{eq:dissfluctsigma}
 \langle\xi(t,\vec x)\xi(t',\vec x')\rangle_\xi=\delta(\vec x-\vec x')\delta(t-t')m_\sigma\eta\coth\left(\frac{m_\sigma}{2T}\right)~.
\end{equation}
To avoid unphysical dependences on the lattice spacing \cite{CassolSeewald:2007ru}, we introduce a correlation of the noise field over a correlation length of $1/m_\sigma$. Hereby, the mass of sigma can be determined as a function of temperature and chemical potential equal to the curvature of the thermodynamic potential in equilibrium, 
\begin{equation}
 \label{eq:corrl}
 m_\sigma^2=\frac{\partial^2 \Omega}{\partial\sigma^2}\bigg|_{\sigma=\langle\sigma\rangle}~.
\end{equation}

The locally equilibrated quark plasma acts as a heat bath in which the field $\sigma$ evolves. The local pressure of this heat bath is given by
\begin{equation}
\label{eq:pressure}
 p(T,\mu; \sigma) = -\Omega_{\rm q\bar q}(T,\mu; \sigma)~,
\end{equation}
allowing us to calculate the local net-baryon and energy densities in the standard fashion as
\begin{equation}
 n=\frac{\partial p}{\partial\mu}~,~~e=T\frac{\partial p}{\partial T}-p+\mu n~.
\end{equation}

As the total energy and momentum of the coupled system of fluid and field are conserved, we obtain the following expressions for the divergences of the ideal energy-momentum tensor of the fluid $T^{\mu\nu}=(e+p)u^{\mu}u^{\nu}-pg^{\mu\nu}$ and the net-baryon current $N^{\mu}=n u^{\mu}$ with the local four-velocity $u^{\mu}$, 
\begin{align}
\label{eq:fluidT}
\partial_\mu T^{\mu\nu}&=-\partial_\mu T_\sigma^{\mu\nu}~,\\
\label{eq:fluidN}
\partial_\mu N_{\rm q}^{\mu}&=0~.
\end{align}
It is worth pointing out that the stochastic nature of the source term on the right hand side of Eq.\ \eqref{eq:fluidT} constitutes a stochastic evolution for the fluid dynamical medium. 
\begin{figure}[t!]
\centering
    \subfloat[\label{fig:varsig}]{
    \centering
    \includegraphics[scale=0.65,angle=270]{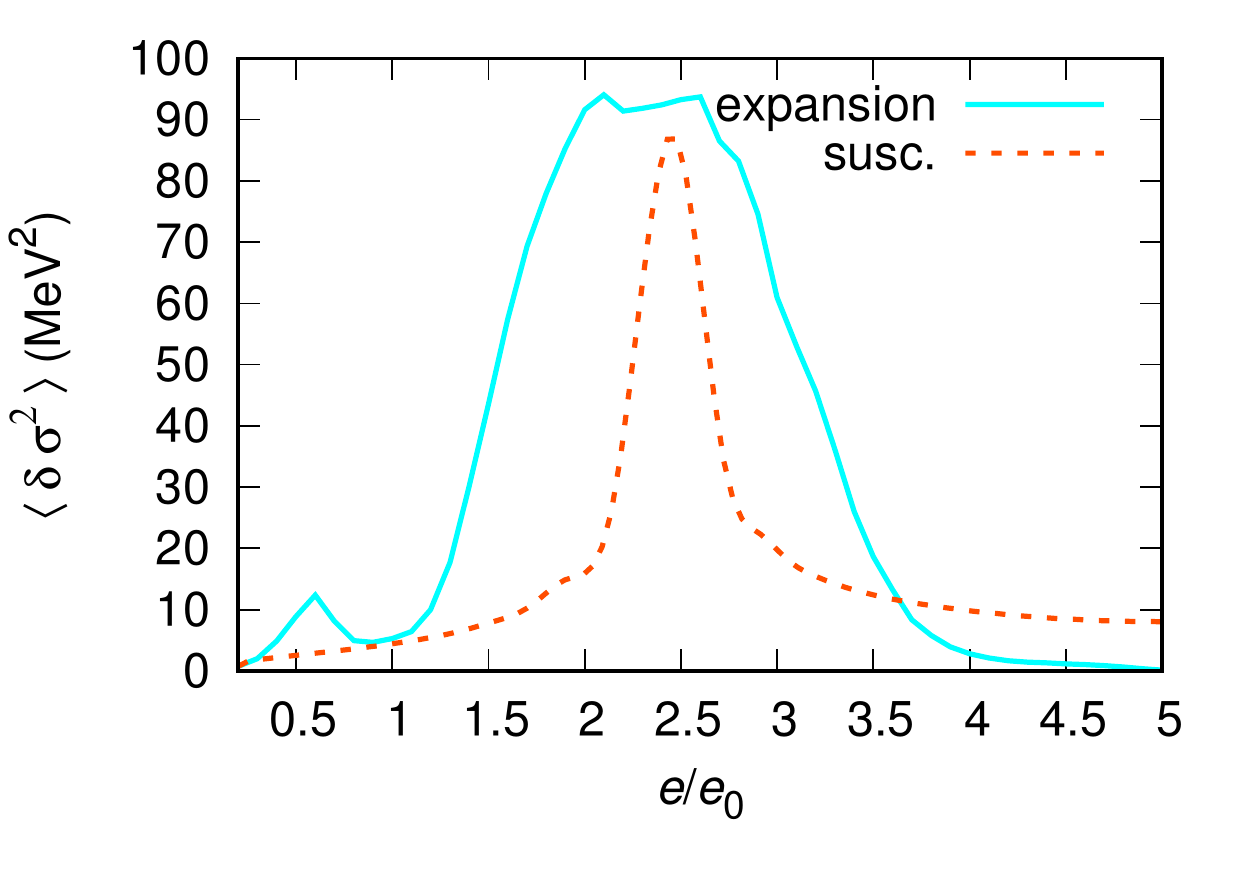}
    }    
  \vfill
    \subfloat[\label{fig:skewsig}]{
    \centering
    \includegraphics[scale=0.65,angle=270]{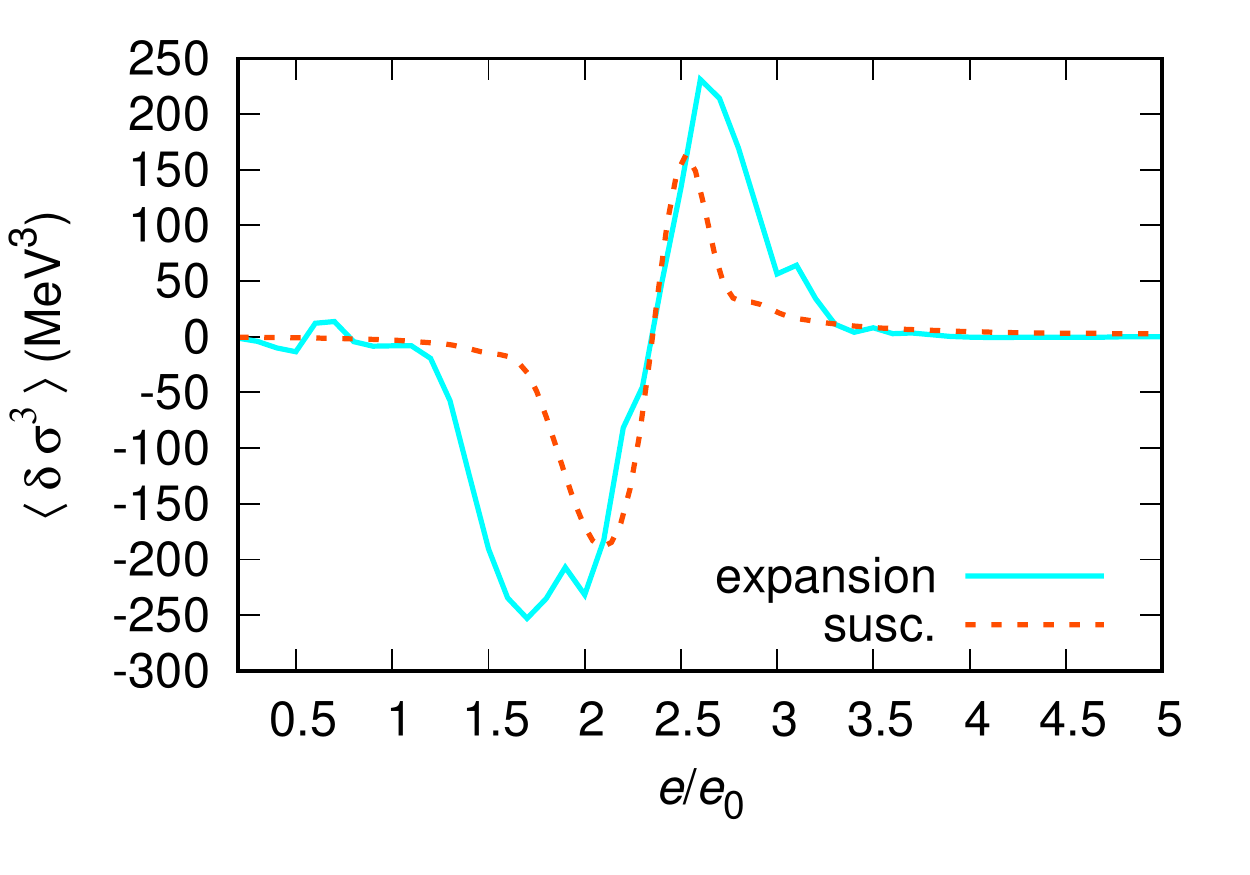}
    }
  \vfill
    \subfloat[\label{fig:kurtsig}]{
    \centering
    \includegraphics[scale=0.65,angle=270]{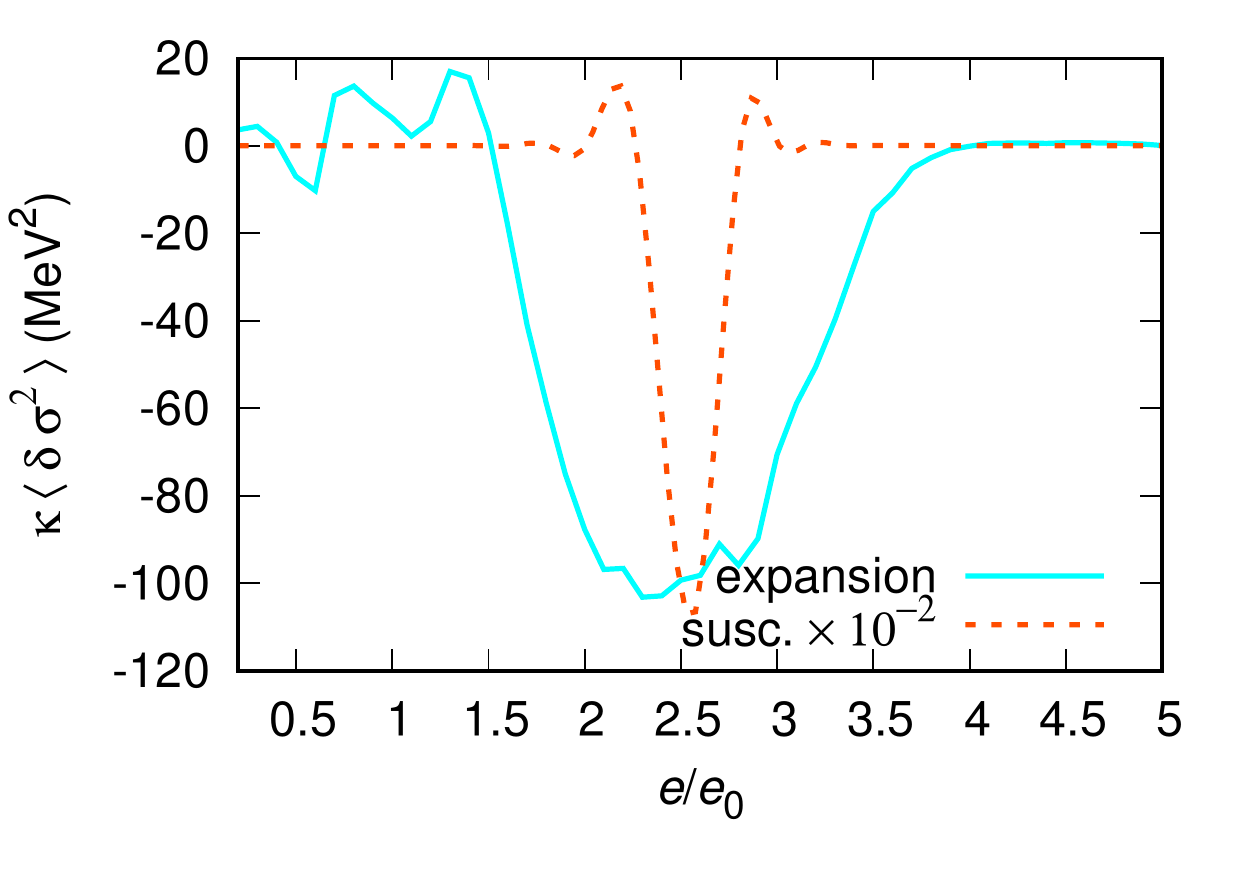}
    }    
\caption[maps]{Second \subref{fig:varsig}, third \subref{fig:skewsig}, and fourth \subref{fig:kurtsig} cumulants of the sigma field for an expanding medium (solid lines) compared to the equilibrium susceptibilities (dashed lines) which have been calculated for the corresponding values of $T$ and $\mu$ along the trajectory of the expanding medium.}
\label{fig:exp}
\end{figure}

\section{Linear cooling in an isothermal box}
\label{sec:box}

We set up a cubic box of $N^3$ grid sites to a global uniform temperature and chemical potential of $T_{\rm ini}=\SI{160}{\MeV}$ and $\mu_{\rm ini}=\SI{100}{\MeV}$, and subsequently let the system evolve according to Eq.\ \eqref{eq:eomsigma} with linearly decreasing the temperature by hand and leaving the chemical potential constant. We use $T(t)=T_{\rm ini}-10\cdot t$, such that after a run time of $t_{\rm run}=\SI{8}{\femto\metre}$, we arrive at a final temperature of $\SI{80}{\MeV}$. This roughly resembles, yet oversimplifies, the situation in a heavy-ion collision at intermediate beam energies. Previously, a roughly linear decrease has been observed in a central volume of the expanding medium using this model \cite{Herold:2013bi}. Within the parameters of the model, the medium in the box then evolves through the crossover transition to the left of the CP. We extract the cumulants $\langle\delta\sigma^2\rangle$, $\langle\delta\sigma^3\rangle$, and $\kappa\langle\delta\sigma^2\rangle$ from the simulation and compare them to the corresponding susceptibilities according to Eqs.~\eqref{eq:sus2}, 
\eqref{eq:sus3}, and \eqref{eq:sus4} as function of the normalized energy density $e/e_0$. From the results in fig.~\ref{fig:box}, we see that the shape of the curves for the dynamical fluctuations follows the shape of the susceptibilities, however, they are shifted to lower energy densities for about $e/e_0=1.0$ to $1.5$ as can be seen from the position of the extrema. Furthermore, the region of criticality, where the cumulants become enhanced, seems clearly extended over for a wider interval in the energy density.

\section{Cooling in an expanding medium}
\label{sec:exp}

In this section, we present results from the evolution of the cumulants of the sigma field in a hydrodynamic expansion realistically resembling the stage of the expanding plasma after the collision of two heavy nuclei in experiment. The initial condition is given by an ellipsoid which is circular in the $x$-$y$ plane with a radius of $6.5$~fm and and has an extension of $4.5$~fm in $z$-direction. Within this shape, an initial temperature and quark chemical potential are given with $T_{\rm ini}=160$~MeV and $\mu_{\rm ini}=130$~MeV. These values are smoothed at the edges by a Woods-Saxon distribution to ensure an analytic transition to the vacuum. From $T_{\rm ini}$, $\mu_{\rm ini}$, the initial values for $\langle\sigma\rangle$, $p$, $e$, $n$ are calculated at each point on the numerical grid. The coupled system of fluid and chiral field sigma is then allowed to evolve according to the equations of motion \eqref{eq:eomsigma}, \eqref{eq:fluidT}, and \eqref{eq:fluidN}. 

As in the previous section, we extract the cumulants $\langle\delta\sigma^2\rangle$, $\langle\delta\sigma^3\rangle$, and $\kappa\langle\delta\sigma^2\rangle$ from the evolution. Here, however, where we are facing an inhomogeneous medium, we cannot simply collect the distributions of $\sigma$ for constant times but rather need to consider them on hypersurfaces of constant energy density, ranging from $e/e_0=0.2$ to $4.0$. The cumulants which are determined from these distributions are compared to susceptibilities according to Eqs.~\eqref{eq:sus2}, 
\eqref{eq:sus3}, and \eqref{eq:sus4}, see fig.~\ref{fig:exp} These susceptibilities are calculated along the trajectories given by the time-dependent volume-averaged values of $T$ and $\mu$ in the central region of the medium. In contrast to our observation from the previous section, where we found a clear delay in the buildup of the characteristic fluctuation signal in all cumulants, here, the enhancement takes place in a region from $e/e_0=1.5$ to $3.5$, compared to enhanced equilibrium fluctuations roughly between $e/e_0=2.0$ and $3.0$. As we have demonstrated previously in \cite{Herold:2016uvv}, the kurtosis of net-protons closely follows the sigma kurtosis during the dynamical evolution at a crossover transition. This has also been argued in \cite{Stephanov:2011pb} for the case of an equilibrated system. The broadening of the critical region can be ascribed to the inhomogeneity of the medium, the temperature and chemical potential on the energy hypersurfaces are not uniquely determined but subject to fluctuations themselves. Therefore, fluctuations in an expanding plasma may build up earlier, at higher energy densities but also remain present for lower energy densities compared to the thermodynamic susceptibilities. This is also strengthened by the influence of memory effects which we have shown to play the dominant role in the case of a homogeneous medium with a uniform temperature and chemical potential. We expect that this behavior increases the chances of measuring observables which are influenced by the critical region of the chiral phase transition.

\section{Summary and Outlook}
\label{sec:sum}

We have studied the behavior of fluctuations in the chiral order parameter sigma both in equilibrium and in a dynamically evolving medium. After deriving a general expression for the second, third, and fourth cumulant of a scalar field, we have applied our results to determine the kurtosis of the sigma field in the quark-meson model. We have seen a characteristic structure with a strong minimum which becomes more enhanced the closer we get to the CP. Then, we have compared these thermodynamic susceptibilities to dynamically evolving cumulants as extracted from the evolution of the sigma field according to a Langevin equation of motion in an isothermal heat bath with linear cooling. Here, a clear delay in the buildup of critical structures is observed. This behavior somewhat changes when we consider a hydrodynamically expanding medium which is self-consistently coupled to the chiral field. This resembles a realistic situation present during the expansion of the hot plasma in a heavy-ion collision. In this setup, the enhancement of cumulants is found in an extended region of energy densities compared to the thermodynamic susceptibilities. The implication is clear: Considering that the net-proton or net-baryon cumulants follow the cumulants of the chiral field, we have an increased possibility to observe interesting nonmonotonic structures in observables of the beam energy scan program or the upcoming FAIR and NICA experiments as well as the BES II program at RHIC. 

In the future, the N$\chi$FD model will be used to study fluctuations and cumulants as function of the chemical potential or beam energy to present an analysis related to the BES program. We will also study the impact of different kinematic cuts.

\section*{Acknowledgements}

The authors are very grateful to Walter Greiner for initiating the collaboration between Goethe University Frankfurt, FIAS and SUT in Thailand which has proven a fruitful source of exchange, ongoing research and productive output over almost 15 years now. 

This work is funded by Suranaree University of Technology (SUT) and by the Office of the Higher Education Commission under NRU project of Thailand. CH and CK acknowledge support from SUT-CHE-NRU (FtR.15/2559) project. This work is further supported by a DAAD exchange grant and the COST Action CA15213, THOR.

The computing resources have been provided by the National e-Science Infrastructure Consortium of Thailand, the Center for Computer Services at SUT and the Frankfurt Center for Scientific Computing.

%
%
\bibliographystyle{unsrt}
\bibliography{mybib}

\end{document}